\documentclass[12pt,tightenlines,floatfix,aps,pra,a4paper,notitlepage]{revtex4-2}

\usepackage{hyperref}

\usepackage{graphicx}
\usepackage{amssymb}
\usepackage{amsmath}
\usepackage{bbm}
\usepackage{dsfont}

\newcommand{\rb}[1]{\left( #1 \right)}
\newcommand{\ew}[1]{\left \langle #1 \right \rangle}
\newcommand{\beq}{\begin{eqnarray}}
	\newcommand{\eeq}{\end{eqnarray}}
\newcommand{\eq}[1]{Eq.~(\ref{#1})}
\newcommand{\fig}[1]{Fig.\,\ref{#1}}
\newcommand{\secref}[1]{Sec.~\ref{#1}}
\newcommand{\bs}[1]{\boldsymbol{#1}}

\renewcommand{\cite}[1]{Ref.~\onlinecite{#1}}

\begin{document}
	
	\title{The persistence of bipartite ecological communities with
		Lotka-Volterra dynamics}
	\author{Matt Dopson \& Clive Emary}
	\affiliation{School of Mathematics, Statistics and Physics,
		Newcastle University,
		Newcastle-upon-Tyne,
		NE1 7RU, UK
	}

	\begin{abstract} 
	The assembly and persistence of ecological communities can be
	understood as the result of the interaction and migration of species. Here we
	study a single community subject to migration from a species pool in which
	inter-specific interactions are organised according to a bipartite network. Considering
	the dynamics of species abundances to be governed by generalised
	Lotka-Volterra equations, we extend work on unipartite networks to we derive
	exact results for the phase diagram of this model. Focusing on antagonistic interactions,
	we describe factors that influence the persistence of the two guilds,
	locate transitions to multiple-attractor and unbounded phases, as well identify
	a region of parameter space in which consumers are essentially absent in the
	local community.
	\end{abstract}

	\maketitle



\section{Introduction}

Understanding patterns in the composition of ecological communities is one of the fundamental goals in ecology \citep{May1972, Berlow1999, McCann2000, Hubbell2001,Wilson2003, Fisher2014}. 
A popular modelling framework for this problem considers in detail a single community in a local habitat embedded within a wider ecosystem  portrayed as a species pool from which the local community can be invaded \citep{MacArthur2001}.
For mathematical analysis, this is then supplemented with several further elements \citep{Wilson2003}.  The first is a dynamical model of the species abundances, and here a generalised Lotka-Volterra approach is typical but other possibilities exist \citep{Lafferty2010,Campbell2011}. The second is a model of interspecific interactions, and in this regard a random-matrix model is often used, having a long history of shedding light on ecological questions  \citep{May1972,May2001} as well as acting as a baseline scenario against which more detailed and ecologically-motivated studies can be compared \citep{Allesina2012}.

Despite the complexity of the resultant community-assembly model, analytical progress has been made \citep{Wilson2003, Servan2018, Pettersson2020}.  In particular, the dynamical cavity method (DCM), a method originating in the physics of disordered systems \citep{mezard1987spin} but since adapted to a number of ecological problems \citep{Rieger1989,Opper1992,tokita2006statistical,yoshino2007statistical,obuchi2016multiple,tikhonov2017collective,advani2018statistical,emary2021can}, has brought significant insight into this model \citep{Bunin2017, Barbier2017, Galla2018}.
\cite{Bunin2017} has given a comprehensive analysis of the phase diagram of this model and show that the DCM solution corresponds to a unique-fixed-point (UFP) phase, in which there exists a unique persistent community that is resistant to invasion.  The DCM solution also gives the boundaries to multiple-attractor (MA), and unbounded phases.

As is typical for the DCM, the model of interactions studied in the above works is statistically homogeneous, i.e. the interaction between all species in the model is described by a single random matrix.  As such, there is no a priori differentiation between the species, and the result is a single abundance distribution for the entire community.
However, we know from the study of ecological networks \citep{Ings2009, Delmas2019, guimaraes2020structure, windsor2023using} that interspecific interactions are anything but homogeneous and that ecological networks possess significant structure, such as trophic levels \citep{Johnson2014}, nestedness \citep{Suweis2013}, or modularity \citep{Grilli2016,Olesen2007}.
One of the most common structures encountered in the network representation of ecological communities is that of the bipartite network, which depicts the interactions between two groups or guilds of species.
The interactions described in these bipartite networks are typically either 
mutualistic, such as in plant-pollinator networks \citep{KaiserBunbury2010, Bane2018, Sheykhali2019}, or 
antagonistic such as in host-parasitoid   
\citep{Cagnolo2010,Morris2013, Hadfield2014, Thierry2019}
or trophic networks 
\citep{Cagnolo2010,Thebault2010,Menke2012,Gilljam2015}.
Bipartite networks are also found as natural components of multi-partite networks, e.g. 
\cite{Pocock2012,Miller2021}, or multitrophic foodwebs e,g, \cite{Williams2011}.

In this paper, we apply the DCM to a species pool and hence local community in which the interactions are structured as a bipartite network but are otherwise random.  We show the applicability of the DCM to this kind of structured scenario and derive analytic results for abundance distributions and persistence probabilities of each of the two guilds.
Focusing on trophic bipartite networks, we describe the phase diagram of the consumer-resource community and show that the UFP, MA and unbounded phases of the unstructured, unipartite model still occur, but with phase boundaries that exhibit non-trivial scaling behaviour.
Furthermore, we report the existence of a region in parameter space in which consumers are effectively absent from the persistent community.


\section{Bipartite community-assembly model}

Our species pool consists of two guilds of species in a bipartite ecological network, i.\,e. with interactions only occurring between species in different guilds. Let $S^{(1)}$ and $S^{(2)}$ be the number of species of each guild in the species pool, $S = S^{(1)}+S^{(2)}$ be the total species number, and $\rho^{(i)} = S^{(i)}/S$ be the corresponding ratios.
We define $N_\alpha^{(i)}$ as the abundance of species $\alpha$ in guild $i$ in the local community, and $r^{(i)}_\alpha$ and $K_\alpha^{(i)}$ as its growth rate and carrying capacity respectively. These letter two quantities we define as positive, with their sign given by coefficient $t^{(i)}\in\{-1,+1\}$ that depends on whether the species in guild $i$ grow or die out in absence of interaction.  We then posit that dynamics of the abundances is described by the generalised Lotka-Volterra equations 
\begin{align}
    \frac{dN_\alpha^{(i)}}{dt} 
    = 
    \frac{r_\alpha^{(i)}}{K_\alpha^{(i)}}N_\alpha^{(i)}
    \left( 
      t^{(i)}K_\alpha^{(i)}-N_\alpha^{(i)}
      +
      c^{(i)}\sum_{\beta=1}^{S^{(i+1)}}a_{\alpha,\beta}^{(i,i+1)}N_\beta^{(i+1)}
    \right)
    \label{EQ:EoM1}
    ,
\end{align}
for species $1\le \alpha \le S^{(i)}$ and guild $i =1,2$ where we adopt a periodic labelling convention that maps $i=3$ onto guild $i=1$.
In \eq{EQ:EoM1}, coefficients $a^{(i,i+1)}_{\alpha,\beta}$ represent the strength of interaction experienced by species $\alpha$ in guild $i$ due to species $\beta$ in guild $i+1$.  We arrange these interaction elements into the $S^{(i)} \times S^{(i+1)}$  matrices $\mathbf{A}^{(i,i+1)}$.
We take all interactions between guild $i$ and $i+1$ to be of the same the type (antagonistic, mutualistic or competitive), and correspondingly set the matrix elements as non-negative $a^{(i,i+1)}_{\alpha,\beta} \ge 0 $, with the signs of the interaction provided by the interguild interaction signs $c^{(i)}\in\{ -1,+1\}$.

We then set the scaling of the matrix elements of $\mathbf{A}^{(i,i+1)}$ with pool size such that 
$\mu^{(i)}:= \sqrt{S}\cdot\langle a^{(i,i+1)}_{\alpha,\beta}\rangle $ and  
$(\sigma^{(i)})^2:=S \cdot\text{Var}(a^{(i,i+1)}_{\alpha,\beta})  $ 
are fixed as $S\to \infty$ with $\rho^{(i)}$ constant. 
This is a natural choice when $a^{(i,i+1)}_{\alpha,\beta}$ are chosen from a non-negative distribution such as the half-normal distribution for which the standard deviation is proportional to the mean [a similar scaling was adopted in \cite{Emary2022}].
We can thus rewrite matrices $\mathbf{A}^{(i,i+1)}$ in terms of the centered, normalised matrix $\mathbf{B}^{(i,i+1)}$ with elements
\begin{align}
    \left\langle b^{(i,i+1)}_{\alpha,\beta}\right\rangle = 0, \qquad\qquad \left \langle \rb{b^{(i,i+1)}_{\alpha,\beta} }^2\right\rangle = 1
    \label{EQ:bmoments}
    .
\end{align}
We express correlations in the different interaction directions by choosing $\mathbf{B}^{(i,i+1)}$ such that they have the property: 
$$
\left \langle b^{(i,i+1)}_{\alpha,\beta} b^{(i+1,i)}_{\beta,\alpha}\right\rangle 
= 
\gamma
.
$$ 
with parameter $\gamma \in [0,1]$.  We restrict ourselves to positive correlations here to avoid any ambiguity in the implied sign assignments of $a^{(i,i+1)}_{\alpha,\beta}$.
In these terms, the interaction blocks become:
\begin{align}
    \mathbf{A}^{(i,i+1)} 
    = 
    S^{-1/2} \left[
    \mu^{(i)}
    \mathbf{J}^{(i,i+1)} 
    + 
    \sigma^{(i)} \mathbf{B}^{(i,i+1)}
    \right]
    \label{EQ:AB}
    ,
\end{align}
where $\mathbf{J}^{(i,i+1)}$ is the $S^{(i)} \times S^{(i+1)}$ matrix of ones.
Without loss of generality, we set the mean carrying capacity to be one,
$\langle K^{(i)}\rangle = 1$, and then further parameterise the carrying-capacities such that 
$
   \rb{\kappa^{(i)}}^2 = S^{(i)}  \cdot \mathrm{Var} \rb{K^{(i)}} 
$ 
is also fixed as $S\to \infty$. This choice is justified a posteriori as being consistent with the interaction scaling.
An alternate to this scaling scheme is discussed later.

In the following, the main quantity of interest will be the fraction of pool species in guild $i$ that persist in equilibrium
\begin{align}
  \phi^{(i)} = 
  \lim_{t\to \infty}\frac{1}{S^{(i)}} 
  \sum_{\alpha=1}^{S^{(i)}} \Theta\rb{N_\alpha^{(i)}}
  ,
\end{align}
in which $\Theta$ is the Heaviside function.

\section{Dynamical cavity method}

For an overview of the DCM in the ecological unipartite context, we refer the reader to the tutorial article of \cite{Barbier2017}, as well as to the work of \cite{Bunin2017}.  
A full account of our derivation of the bipartite case is given in Appendix~\ref{SEC:DCM}, but the essence of the method is that an equilibrium configuration is considered to which a new species from each guild is added.  The action of the pre-existing community on the added species is treated exactly, but the reciprocal action of the added species on the community is small in the large-$S$ limit and treated in linear response.  Since the added species are identical with other species from the same guild, this leads to a closed system of equations that can be solved  self-consistently.  A discussion of our solution technique for the equations is given in Appendix\,\ref{SEC:Asol} and a discussion of the validity given in Appendix\,\ref{SEC:accuracy}.

The central result in this analysis is that the abundances of species within a guild are each distributed according to truncated Gaussians \citep{Wilson2003,Bunin2017,Servan2018,Galla2018,Pettersson2020}
with interdependent parameters. The key properties of the distributions are described by two quantities $\Delta^{(i)}; i=1,2$.  In particular, the fraction of species in guild $i$ that persist in equilibrium is given by
\begin{align}
    \phi^{(i)} = w_0(\Delta^{(i)})
    ,
\end{align}
in which $w_k$ are a set of functions defined via
\begin{align}
    w_k(\Delta^{(i)}) := \int_{-\Delta^{(i)}}^\infty (\Delta^{(i)}+z)^k\frac{1}{\sqrt{2\pi}}e^{-z^2/2}dz 
    \label{EQ:wDef}
    .
\end{align}
In Appendix\,\ref{SEC:DCM} we give details of the equations that determine  $\Delta^{(i)}$ in the most general case. Here we just reproduce them in the simplest $\gamma = \kappa^{(1)} = \kappa^{(2)} = 0$ case:
\begin{align}
    \rho^{(1)} \rho^{(2)} \rb{\sigma^{(1)} \sigma^{(2)}}^2  w_2(\Delta^{(1)})w_2(\Delta^{(2)})
    =
    1
    \label{EQ:singeqn1main}
    ,
\end{align}
and
\begin{align}
    t^{(2)}\sigma^{(1)}\Delta^{(1)}
    \sqrt{\rho^{(2)} w_2 (\Delta^{(2)})}
    -
    c^{(1)}  t^{(2)}\sqrt{S}\rho^{(2)}\mu^{(1)} w_1(\Delta^{(2)})
    ~~~~~~~~~~~~~~~~
    \nonumber\\~~~~~~~
    =    
     t^{(1)} \Delta^{(2)} 
    -
    t^{(1)} c^{(2)}  \sigma^{(1)} \mu^{(2)}\rho^{(1)} w_1(\Delta^{(1)}) 
    \sqrt{S\rho^{(2)} w_2 (\Delta^{(2)})}
    \label{EQ:singeqn2main}
    .
\end{align}

\subsection{Phase diagram}

Our main focus will be on antagonistic interactions and for concreteness we will use the language of trophic interactions  In this setting, we identify guild 1 with the resource species and guild 2 with consumers. The corresponding choices of sign are
$
  t^{(1)} = - t^{(2)} = +1
$,
such that in the absence of interactions, the resource-species abundances grow to carrying capacity and the consumers die out; and
$
  c^{(1)} = - c^{(2)} = - 1
$
such that the interaction is beneficial to the consumers and detrimental to the resources.

As in the unipartite case, the bipartite model is found to exhibit three phases: the UFP described by the cavity solution outlined above, plus the unbounded and MA phases.
In the unbounded phase, one or more of the species abundances diverge, such that one or both of $ \lim_{t\to\infty}\ew{N^{(i)}} \to \infty$.
For the abundances in the cavity solution to remain bounded, we require that both sides of \eq{EQ:singeqn2main} (in the $\kappa^{(i)}=\gamma =0$ case) are greater than zero.  As shown in Appendix~\ref{SEC:stability}, the implication of this is that the location of phase boundary is asymptotically given by 
\begin{align}
    \frac{1}{2} \rho^{(1)} \rho^{(2)} \rb{\sigma^{(1)} \sigma^{(2)}}^2
    w_2\rb{ - \frac{\sqrt{S_2} \mu^{(1)}}{\sqrt{\pi}\sigma^{(1)}} }
    = 1
    ,
\end{align}
which holds for $\gamma = 0$ but arbitrary $\kappa^{(i)}$.
As example, let us assume that the interaction strengths are distributed according to a half-normal distribution for which mean and standard deviation are related as $\mu^{(i)} = \sigma^{(i)} \sqrt{2/(\pi - 2)}$. Then, assuming that the couplings are symmetric, $\sigma^{(1)} = \sigma^{(2)} = \sigma$,  we find that the unbounded phase occurs when $\sigma > \sigma_c$ with critical interaction strength 
\beq
  \sigma_c =
   \left\{
     \frac{2}{ 
      \rho^{(1)} \rho^{(2)}
      w_2\left[ - \sqrt{2S\rho^{(2)}/(\pi(\pi-2))}\right]
     }
  \right\}^{1/4}
  \label{EQ:sigmac}
  .
\eeq
The key observation is that because
$
  w_2(-|\Delta|) \sim
  \sqrt{\frac{2}{\pi}} e^{-\frac{1}{2}\Delta^2}  /|\Delta|^3
$ 
for large $|\Delta|$, the critical interaction strength $\sigma_c$ diverges exponentially with pool size $S$ and 
becomes inaccessible. At finite $S$, the transition occurs at finite interaction strength.

Stability analysis of the cavity solution shows that, as in the unipartite case, it becomes unstable and gives way to an MA phase.
Appendix~\ref{SEC:MA} shows that the boundary to the MA phase obeys the equation (valid for $\gamma = 0$, arbitrary $\kappa^{(i)}$)
\begin{align}
    \rho^{(1)} \rho^{(2)} \rb{\sigma^{(1)} \sigma^{(2)}}^2  w_0(\Delta^{(1)})w_0(\Delta^{(2)})
    =
    1.
\end{align}
To find the parameters of this boundary in the $\kappa^{(i)} =0$ case, we look for overlap of this curve with that described by \eq{EQ:singeqn1main}.  Since both curves as symmetric with respect to interchange of $\Delta^{(i)}$, the boundary behaviour where the two curves just cease to overlap occurs when $\Delta^{(1)} = \Delta^{(2)} = \Delta$.  From this, we determine that the critical parameters occur when $ w_0(\Delta) = w_2(\Delta)$.  This happens at $\Delta = 0$, at which point $w_0(\Delta) = \frac{1}{2}$. The result is that the critical parameters for the MA transition obey
\begin{align}
    \rho^{(1)} \rho^{(2)} \rb{\sigma^{(1)} \sigma^{(2)}}^2  
    =
    4.
\end{align}
For $\sigma^{(1)}=\sigma^{(2)}$, the MA phase therefore occurs when
$
  \sigma > \sigma_\mathrm{MA} = \left[4/(\rho^{(1)}\rho^{(2)})\right]^{1/4}
$.

\section{Results}

\begin{figure}[tb]
\centering
 \includegraphics[width=0.8\columnwidth]{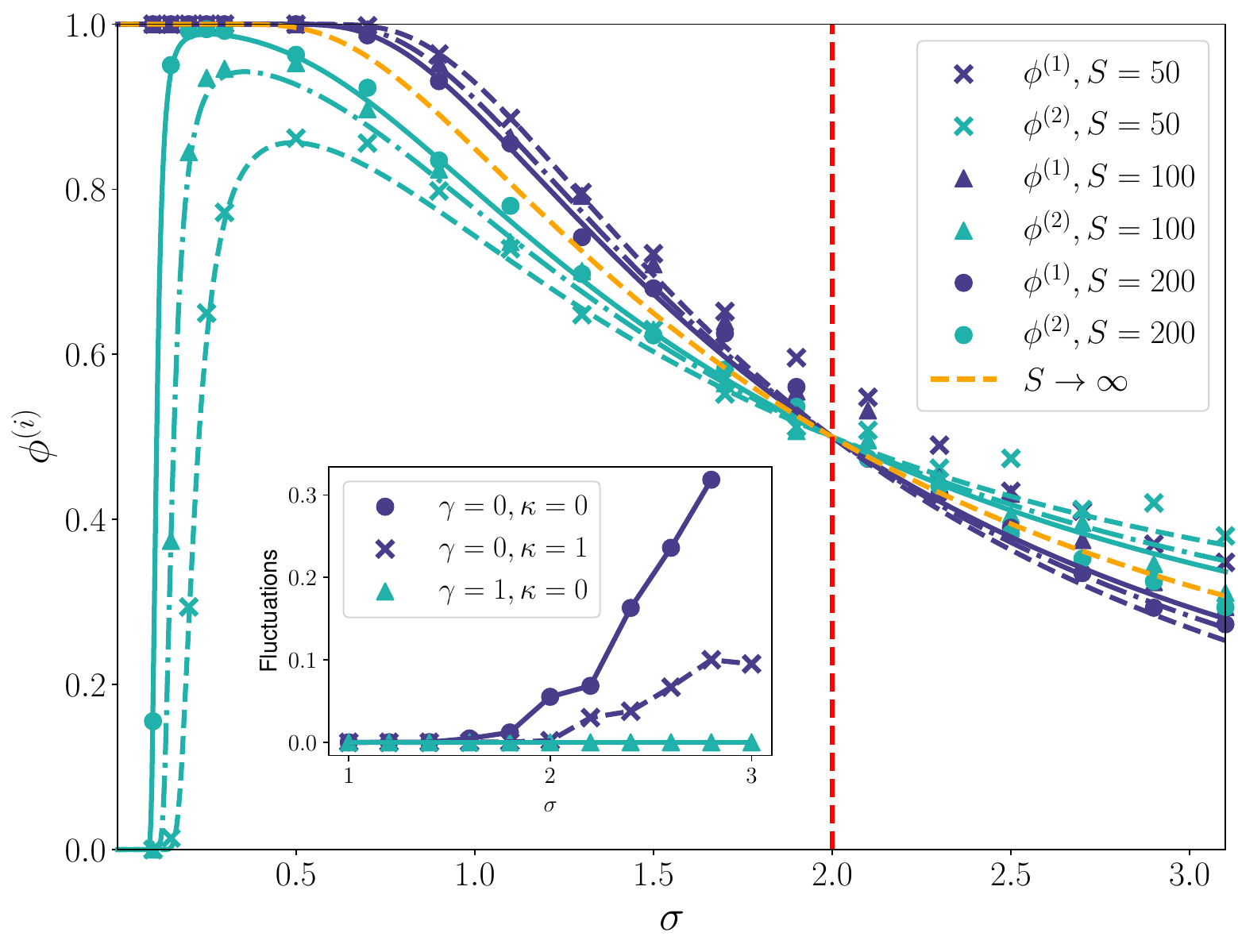}
  \caption{
    The fraction of persistent resources  ($\phi^{(1)}$, dark blue) and consumers ($\phi^{(2)}$, light blue) as a function of interaction strength $\sigma$ with parameters $\kappa^{(i)} = \gamma = 0$, $\rho^{(i)} = \frac{1}{2}$, and with interactions symmetric, $\sigma^{(i)} = \sigma$, and drawn from a half-normal distribution.
    Good agreement is seen between analytic results (lines) and numerical simulations (markers) and this increases as the size of the species pool $S$. The red dashed line indicates the transition from a unique fixed point (UFP) to multiple-attractor (MA) phase. For small interaction strengths, the consumer fraction drops to almost zero. 
    The orange line shows the asymptotic $S \to \infty$ limit, in which both $\phi^{(i)}$ are identical for these parameters.
    INSET: Relative fluctuations in equilibrium abundances of the numerical solution taken over realisations of the interaction matrix and carrying capacities. Marked increases occur around the critical interaction strength of the MA transition: $\sigma = 2$ for $\gamma=\kappa = 0$; $\sigma = 2.39$ for ($\gamma=0$, $\kappa = 1$). For ($\gamma=1$, $\kappa = 0$), the transition is outside this $\sigma$-range, and no increase in fluctuations in seen. 
  }
  \label{FIG:GLVvsDCM_00}
\end{figure}

In visualising the results of this calculations we reduce the number of independent parameters by choosing $\sigma^{(1)} = \sigma^{(2)} =\sigma$, and $\kappa^{(1)} =\kappa^{(2)} = \kappa$. Furthermore, although $\mu^{(i)}$ and $\sigma^{(i)}$ are in general independent parameters, here we consider them to be derived from a half-normal distribution for which they are related as $\mu^{(i)} = \sigma^{(i)} \sqrt{2/(\pi - 2)}$.

In \fig{FIG:GLVvsDCM_00} we plot the persistent fractions $\phi^{(i)}$ as a function of interaction strength $\sigma$ in the simplest case of $\kappa = \gamma =0$. 
We show analytic results from the DCM described previously as well as results obtained from numerical simulations of the GLV equations (described in Appendix~\ref{SEC:numerics}).  Overall agreement is good, and is seem to improve for larger values of the pool size $S$. As in the unipartite case, the DCM solution still gives a good account of the simulation results in the MA phase (to the right of the red dotted line in \fig{FIG:GLVvsDCM_00}), despite the loss of stability of the cavity solution in this region.

\begin{figure}[tb]
\centering
 \includegraphics[width=0.8\columnwidth]{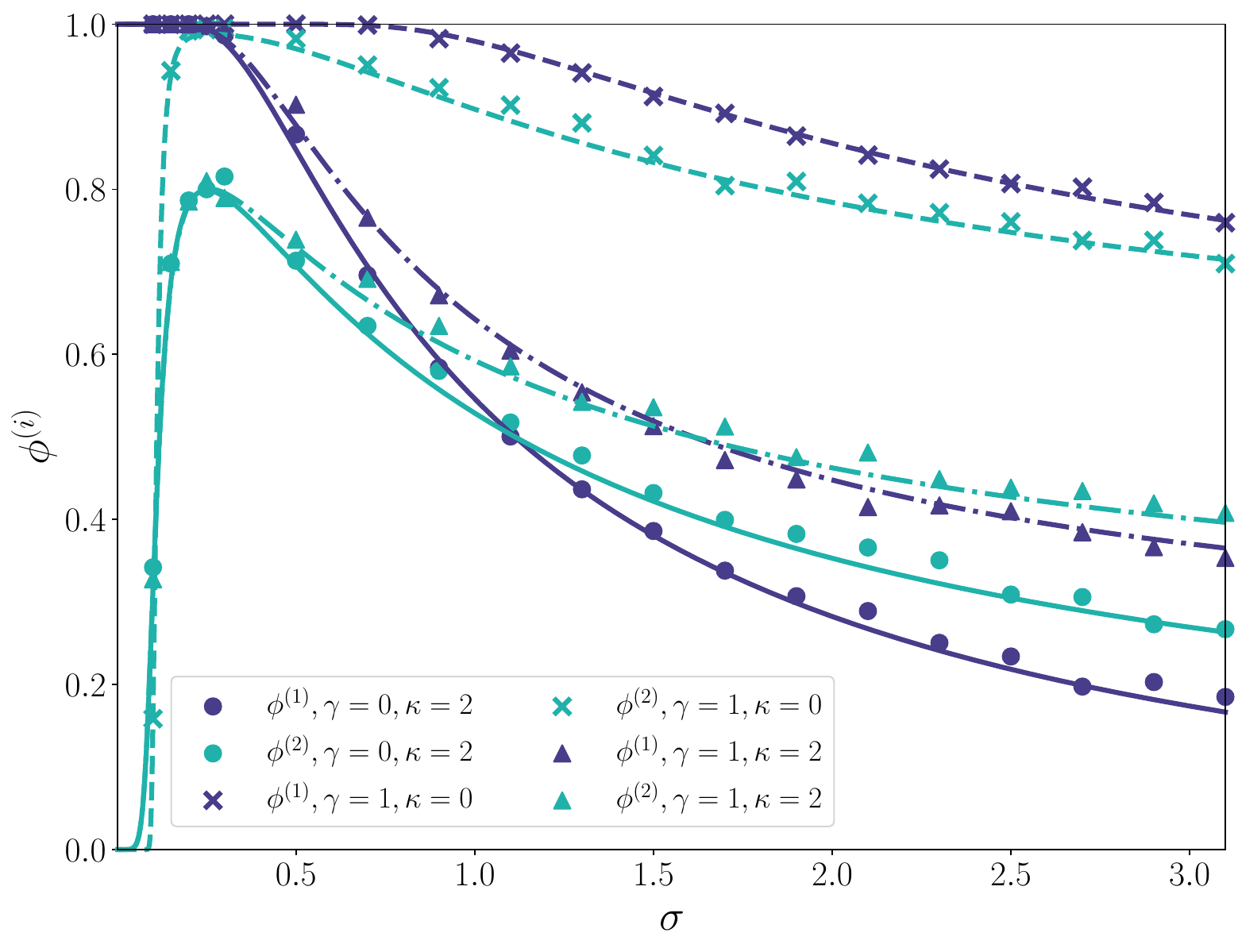}
  \caption{
    As \fig{FIG:GLVvsDCM_00} but with different $\kappa^{(i)} = \kappa$ and $\gamma$ assignments. The dark blue lines are for $\phi^{(1)}$ and light blue are for $\phi^{(2)}$, with line and marker type indicated different values of $\kappa$ and $\gamma$ parameters.  We once again see good agreement between analytics and simulations. Other parameters were: $S=200$, $\rho^{(1)}=\rho^{(2)} = \frac{1}{2}$
}
  \label{FIG:GLVvsDCM_mixed}
\end{figure}

Figure~\ref{FIG:GLVvsDCM_mixed} shows that this agreement between simulations and analytics also extends to non-zero values of $\kappa$ and $\gamma$, and thus that the DCM is able to address correlations in the interaction matrix and a distribution of carrying capacities in this structured context.

The inset of \fig{FIG:GLVvsDCM_00} gives numerical evidence of the location of the MA transition. Here we plot the standard deviation of the abundance values $N_\alpha^{(i)}$ sampled over initial conditions, and averaged over instances of interaction matrix and carrying capacities and over all species in both guilds. For presentation, this measure of the fluctuations in the final equilibrium state is normalised against the mean abundance over all runs. We see that, for an interaction strength lower than the critical value predicted by the DCM, the fluctuations are extremely small, but that around the critical interaction strength they start to rise.  Clearly this is indicative of the unique fixed point below the transition giving way to the MA phase above it, as the multiple attractors will have different abundance distributions and hence finite fluctuations from instance to instance.

From Figs.\,\ref{FIG:GLVvsDCM_00} and \ref{FIG:GLVvsDCM_mixed} we see that, even though the parameters are chosen symmetrically, there are significant differences between the two persistence fractions. In particular, at smaller interaction strengths, $\phi^{(1)}$ (resources) is larger than $\phi^{(2)}$ (consumers) and at larger interaction, this trend is reversed. This is perhaps not surprising, given the different roles of the two guilds. However, in the asymptotic limit ($S \to \infty$; see Appendix~\ref{SEC:Aasym}) for  $\kappa=\gamma =0$ and symmetric parameters we find that $\Delta^{(1)} = \Delta^{(2)} = \Delta$ with $\Delta$ determined by
$
    w_2(\Delta) = 2/\sigma^2
$ for any $\sigma \ne 0$. In this case, then, the abundance distribution of the two guilds becomes identical and the two $\phi^{(i)}$ curves overlap (a result shown as an orange line in \fig{FIG:GLVvsDCM_00}).

Apparent from these plots is that for a range of interaction strength from zero upwards, the persistence fraction of the consumers is suppressed to being close to zero, such that at these values of interaction strength consumers will be effectively absent from the community.  Analytic insight into this region can be obtained as follows. 
We assume that $\Delta^{(1)} \gg 1$ (consistent with $\phi^{(1)} \approx 1$) and correspondingly approximate $w_k(\Delta^{(1)}) \approx \rb{\Delta^{(1)}}^k$. This together with \eq{EQ:singeqn1main} allows the elimination of $\Delta^{(1)}$ from \eq{EQ:singeqn2main}. Since $\Delta^{(2)} \ll -1$ (consistent with $\phi^{(2)} \approx 0$) we then can approximate  $w_k(\Delta^{(2)})\approx 0$ in \eq{EQ:singeqn2main}, which yields an equation for $\Delta^{(2)}$ in the suppressed region.  Generalising this to the  $\kappa^{(i)}\ne0$ case, the result we obtain is
\beq
  \Delta_2 
  \approx
  \frac{
    \sqrt{S} \rho^{(1)}\mu^{(2)} -1
  }{
   \sqrt{ \rho^{(1)} \rb{\sigma^{(2)}}^2 + \rb{\kappa^{(2)}}^2 (S^{(1)})^{-1}}
  }. 
\eeq
To be compatible with $\phi^{(2)} \ll 1$ this needs to large and negative, which requires $ \sqrt{S} \rho^{(1)}\mu^{(2)} \ll 1$. This suggests that 
\beq
  \mu^{(2)}  = \frac{1}{\rho^{(1)}\sqrt{S} } 
  \label{EQ:suppressionline}
\eeq
(at which point $\Delta_2 \approx 0 $ and hence $\phi^{(2)} = w_0(0) = \frac{1}{2}$) gives an indicator of when the consumer population rises to a significant value. Below this point, the 
 suppression of the consumers is exponential because for $\Delta_2$ is large and negative, we can approximate
$
  \phi^{(2)} = w_0 \rb{-|\Delta^{(2)}|} \sim e^{- (\Delta^{(2)})^2/2}/(|\Delta^{(2)}|\sqrt{2\pi})
$.
From \eq{EQ:suppressionline}, we see that the width of region without consumers reduces  as the size of the species pool $S$ increases.

\begin{figure}[p]
\centering
  
 \includegraphics[width=1\columnwidth,clip=true]{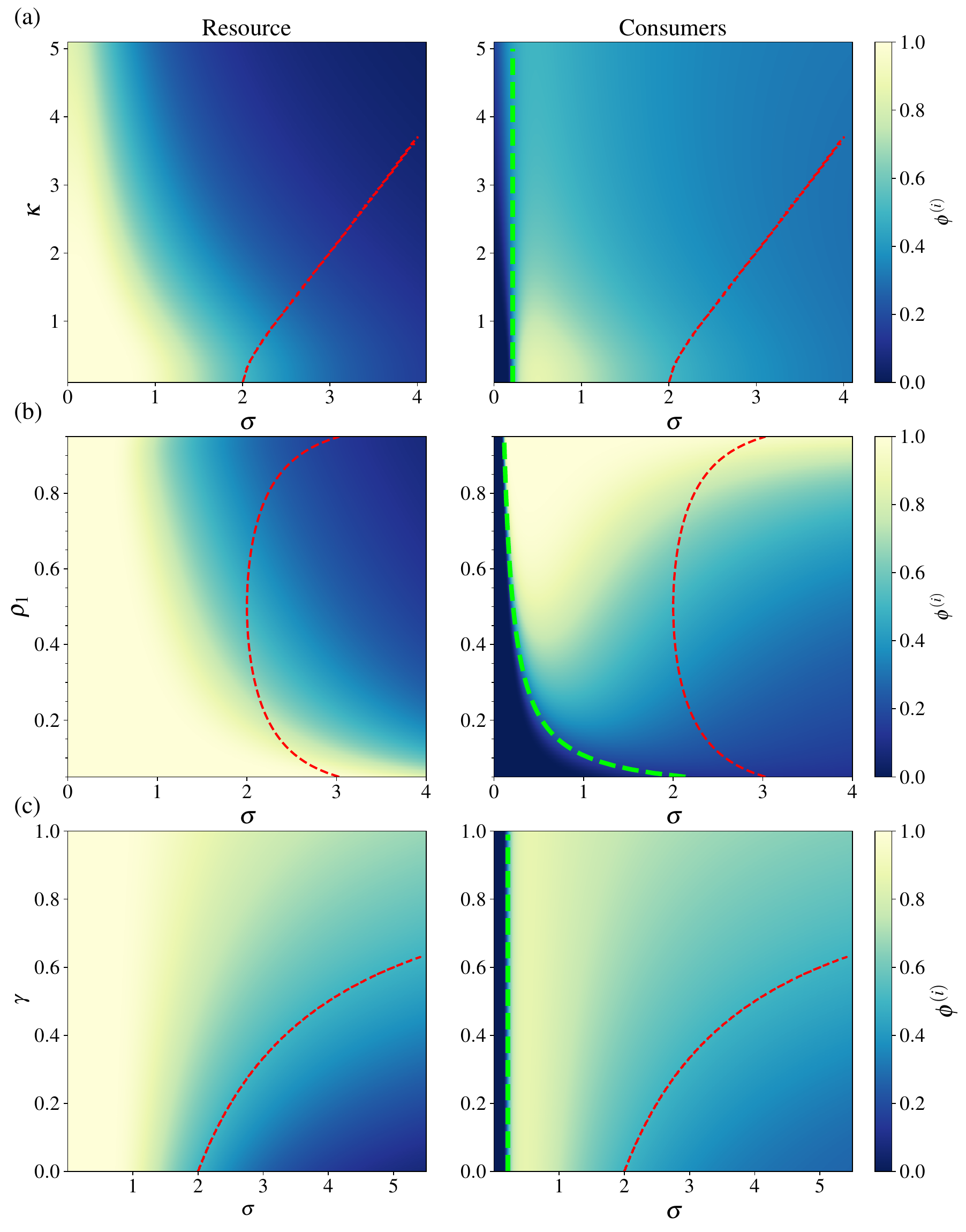}
  \caption{
    Phase diagrams for the antagonistic bipartite networks. The colour scale represents the persistent fraction of resource species ($\phi^{(1)}$, left) and consumers ($\phi_2$, right).
    \textbf{(a)} $\phi^{(i)}$ as a function of $\sigma$ and $\kappa$ with fixed
    $\gamma=0$ and $\rho^{(1)}=1/2$; 
    \textbf{(b)} $\phi^{(i)}$ as a function of $\rho^{(1)}$ and $\sigma$ with 
    $\kappa = \gamma = 0$;
    \textbf{(c)}  $\phi^{(i)}$ as a function of  $\gamma$  and $\sigma$ with 
    $\kappa = 0$ and $\rho^{(1)}=1/2$.
    The red dashed lines show the transition from unique fixed point (UFP) to multiple-attractors (MA) phase. The green dashed lines show the boundary of the consumer suppression region, \eq{EQ:suppressionline}.
    Parameters were $S=50$, $\sigma^{(1)}=\sigma^{(2)}=\sigma$, $\mu^{(1)} = \mu^{(2)} = \sigma \sqrt{2/(\pi-2)}$ and $\kappa^{(1)}=\kappa^{(2)}=\kappa$, for all plots. 
  }
  \label{FIG:phaseplots}
\end{figure}

Figure~\ref{FIG:phaseplots} show different aspects of the phase diagram of the antagonistic bipartite model.  We plot results for a pool size of $S=50$, such that the consumer-suppression regions are easily visible on the same scale as the MA transition. Note that we do not show that unbounded transitions as, according to the arguments presented in the previous section, these occur as large values of interaction strength.

Figure\,\ref{FIG:phaseplots} shows the persistent fractions $\phi^{(i)}$ as a function of interaction strength $\sigma$ and carrying capacity width $\kappa$.  Increasing $\kappa$ results in the persistent fractions dropping more quickly as $\sigma$ increases, and also a noticeable drop in the peak number of consumers. This is a result of the wider distribution of $K_\alpha^{(i)}$ values giving resource species a carrying capacity closer to zero and therefore more likely to becomes locally extinct and consumers (unsigned) carrying capacities further from zero, and therefore more likely to die out rapidly. The red dashed line shows the point of the MA transition, which is seen to move to higher interaction strength with increasing $\kappa$. Finally, the green dashed line shows the boundary \eq{EQ:suppressionline} of the suppressed-consumer region. This is a constant as a function of  $\kappa$ here, but as $\kappa$ increases, the transition out of the suppression region becomes less sharp.

Figure~\ref{FIG:phaseplots}(b) shows how the persistent fractions change as a function of $\sigma$ and $\rho^{(1)}$, the fraction of resource species in the species pool.
In this case we see how the consumer-suppression region depends on the composition of the species pool with the suppression becomes more extensive for smaller $\rho^{(1)}$, i.e. fewer resource species in the pool. The opposite effect is also observed   --- for large values of the resource fraction  $\rho^{(1)}$, the consumer shows an extensive range of interaction strength for which $\phi^{(2)} \approx 1 $ and thus all consumers in the pool are supported in the local community.  The MA transition line is symmetric about $\rho^{(1)} = 1/2$, given the dependence on $\rho^{(1)}\rho^{(2)} = \rho^{(1)}(1-\rho^{(1)})$.

Finally, \fig{FIG:phaseplots}(c) looks at the role of the correlation parameter $\gamma$ in determining community persistence. Increasing correlations ($\gamma>0$) universally serves to increase both persistent fractions, with extensive regions where $\phi^{(i)} > 0.5$ for both consumers and resources.  Transition to the MA phase is moved to higher $\sigma$ for increasing $\gamma$ and being in the MA phase is correlated with a drop in both persistent fraction  values.

\section{Discussion}

We have shown here that the DCM generalises to structured ecological models, and specifically to a bipartite structure with consumer-resource Lotka-Volterra dynamics. We have seen that this model exhibits phases analogous to those of the unipartite model, and that the DCM allows us to map the boundaries between them.
The key feature of the bipartite model is the existence of two guilds, and we have seen that the composition of the persistent community in terms of these two guilds depends both on the strength of the interaction between them, the guild size in the species pool, as well as parameters such as $\gamma$ and $\kappa^{(i)}$.  Interestingly,  in the $S \to \infty$ limit the ratio of the mean equilibrium abundances becomes 
\begin{align}
  \mathcal{R} 
  :=
  \frac{\sum_{\alpha=1}^{S^{(2)}} N ^{(2)}_\alpha}
  {\sum_{\alpha=1}^{S^{(2)}} N ^{(1)}_\alpha}
  = 
  \frac{
     \mu^{(2)} 
    }{
     \mu^{(1)} 
    }
    \label{EQ:Rasymsimple}
  ,
\end{align}
which depends only on the ratio of mean interaction strengths. The ratio of the persistent fraction is not as simple.

One novel feature of our results is that, for weak interactions, the fraction of persistent consumers, $\phi^{(2)}$, is suppressed.   This is ecologically reasonable, as it means that if trophic interactions are too weak, consumers can not be sustained in the community.  What is perhaps surprising is that the transition to a sustained consumer presence is reasonably abrupt in parameter space.  Furthermore, we have shown that the width of this suppressed region depends on the size of the species-pool, becoming narrower as $S$ increases.  For interaction strengths above the consumer onset, properties of the two guilds become similar with both $\phi^{(i)}$ falling off with interaction strength.

The most obvious baseline for comparison is the unipartite model with $\gamma = -1$, as this has interactions arranged in consumer-resource pairs \footnote{Note that the significance of the $\gamma$ parameter is slightly different in the unipartite model because in the bipartite model we have factored out the $c^{(i)}$ signs}. The persistent fraction in that model shows a monotonic decrease with interaction strength, similar to the asymptotic behaviour of the bipartite model in \fig{FIG:GLVvsDCM_00}.  The unipartite model shows no suppression for small couplings because, although the interactions are all trophic, their orientations are random. In contrast, the bipartite structure enforces a consistent direction to the interactions and this paves the way for guild-level effects.

Whilst we have focused on trophic interactions, different choices of the sign factors $t^{(i)}$ and $c^{(i)}$ allow different interguild interactions to be studied.
With $c^{(i)} = +1$ we have a mutualistic bipartite network, either obligate ($t^{(i)}=-1$) or facultative ($t^{(i)}=+1$). Adapting the reasoning from the trophic case, we find that, irrespective of $t^{(i)}$, the transition to the unbounded phase in this mutualistic model occurs for large $\Delta^{(i)}$, and this allows to obtain the transition point as 
$
    \mu^{(1)}\mu^{(2)}
  = \rb{\rho^{(1)}\rho^{(2)}S}^{-1}
$.
Taking the case of symmetric coupling drawn from the half-normal distribution, the critical coupling is given by 
\beq
  \rb{\sigma^\mathrm{MUT}_c}^2 = \frac{\pi(\pi-2)}{2\rho^{(1)}\rho^{(2)}S}
  ,
\eeq
such that bounded phase is obtained when $\sigma < \sigma_c \sim S^{-1/2}$.  The extent of this bounded phase therefore decreases with pool size and vanishes in the limit. And thus the characteristic behaviour of mutualistic bipartite interactions is towards non-persistence of the community.
This might be reconciled with the manifold observation of bipartite mutualistic networks in nature in a number of ways. It might imply that mutualistic interactions are very weak, but this seems unlikely given the important role these interactions typically play in the lifecycles of the participants. It could also indicate limitations in the dynamical model --- inclusion of saturating interactions is an obvious improvement that could be made.  But perhaps the most interesting possibility is that the model suffers from studying mutualism in isolation, and in nature these mutualistic networks modules exist as modules in larger networks with interactions of various types. Finding the networks conditions which allows mutualisms to persist therefore becomes an important future question.

The sign allocation $t^{i)} = +1$ and $c^{(i)} = -1$ gives a model in which the two guilds compete with one another.  In this case the transition to unboundedness occurs when
\beq
  \frac{\mu^{(1)}\mu^{(2)}}{\rb{\sigma^{(1)}\sigma^{(2)}}^2}
  =\frac{4}{S}
  .
\eeq
With symmetric couplings and matrix elements from the half-normal distribution again, this translates into a critical coupling
\beq
 \rb{ \sigma^\mathrm{COMP}_c}^2 = \frac{S}{2(\pi-2)}
  ,
\eeq
such that the bounded phase occurs for $\sigma < \sigma_c \sim S^{1/2}$. The two guilds therefore both always persist in the large-pool limit.
Although we were unable to find reports of bipartite competition networks in the literature (presumably due to the difficulty of observing such interactions), ``negative non-trophic'' interactions have been reported as part of larger multi-interaction networks \citep{Kefi2015,Kefi2016} and often with a particular association with facilitation \citep{Losapio2021}.
In both mutualistic and competitive cases, the symmetry in interaction sign,  $c^{(1)}=c^{(2)}$, ensures that neither guild is significantly suppressed relative to the other.

It is interesting to compare these results for the persistence of communities with different interaction types with the conclusions derived from a linear stability analysis of the relevant interaction matrices
\beq
  \mathbf{A} =
  \begin{bmatrix}
    -\mathbbm{1} &\mathbf{A}^{(1,2)}\\\mathbf{A}^{(2,1)}&-\mathbbm{1}
  \end{bmatrix}
  \label{EQ:Apmatrix}
  ,
\eeq
in which $\mathbbm{1}$ is a unit matrix, and where we set $\gamma = 0$ for simplicity such that $\mathbf{A}^{(1,2)}$ and $\mathbf{A}^{(2,1)}$ are independent.
From e.g. \cite{Emary2022} and references therein, we know that the asymptotic spectrum of $\mathbf{A}$ will consist of two parts: a bulk, and a pair of isolated ``macroscopic'' eigenvalues. In the uncorrelated case, the bulk spectrum will be a circle in the complex plane \citep{Tikhomirov2011} with centre at (-1,0) and radius of  
$
  \sqrt{\sigma^{(1)} \sigma^{(2)}} \rb{\rho^{(1)}\rho^{(2)}}^{1/4}
$.
The macroscopic eigenvalues are given by
\begin{align}
  \lambda_\mathrm{macro,\pm} = -1 \pm
  \left\{
     c^{(1)} c^{(2)} \mu^{(1)} \mu^{(2)} \rho^{(1)} \rho^{(2)} S
  \right\}^{1/2}
  ,
\end{align}
with the scaling $\sim \sqrt{S}$ justifying the ``macroscopic'' moniker.
In the consumer-resource case, $ c^{(1)} c^{(2)} = -1$, and the macroscopic contribution to the spectrum is purely imaginary. Thus it is the bulk that determines the stability. 
In contrast, for both competitive and mutualistic interactions, we have  $ c^{(1)} c^{(2)} = 1$ and the stability properties of these two interaction types will be the same. In these cases we have $\lambda_\mathrm{macro,+}$ real and positive and therefore this eigenvalue dominates stability considerations.  The trend, then, from local stability analysis is that both competitive and mutualistic bipartite interactions are unstable, whereas antagonistic interactions are stable. This stands in contrast with the DCM results which identifies the antagonistic and competitive structures as persisting, whilst the mutualistic one does not.

We now discuss the scaling of the interaction coefficients \citep{Dougoud2018}. The most obvious alternative to \eq{EQ:AB} is to set
\begin{align}   
    \mathbf{A}^{(i,i+1)} 
    = 
    \frac{\mu^{(i)}}{S}
    \mathbf{J}^{(i,i+1)} 
    + 
    \frac{\sigma^{(i)}}{\sqrt{S}} \mathbf{B}^{(i,i+1)}
    \label{EQ:ABoverS}
    ,
\end{align}
so that the mean scales like the variance (rather than the standard deviation). This scaling is like that in the unipartite model \citep{Bunin2017}.  The difference between these two choices is perhaps best appreciated from the spectrum of $\mathbf{A}$ of \eq{EQ:Apmatrix}.  In the alternative scaling, $\mathbf{A}$ has the same bulk spectrum as before but now the $S$-dependence of macroscopic eigenvalues removed. Thus, eigenvalues $\lambda_\mathrm{macro,\pm}$ cease to be macroscopic, and the entire spectrum scales as $S^0$.
The DCM equations for this alternative scaling can be obtained from those presented here by scaling $\mu^{(i)}\to\mu^{(i)}S^{-1/2}$ and $\kappa^{(i)} \to \kappa^{(i)}\sqrt{S^{(i)}}$.  This results in a modification of the phase diagram.  Confining ourselves to the trophic case, the transition to the unbounded phase now  occurs at coupling strength $\sim S^0$, rather than $\sim S^{1/2}$, and the width of the suppressed zone becomes $\sim S^0$ rather than $\sim S^{-1/2}$.  On the other hand, the MA transition remains in the same place, being determined by $\sigma^{(i)}$ and not $\mu^{(i)}$.
Distributions that scale like \eq{EQ:AB} are straightforward to realise -- the half-normal-distribution used here is a simple example.  Not so for \eq{EQ:ABoverS} as this requires a distribution defined on non-negative support $a>0$ in which the ratio of mean to standard deviation scales like $S^{-1/2}$, this inevitably results in a heavily skewed distribution\footnote{An example model that realises this is a Bernoulli distribution in which the probability the matrix element is non-zero is proportional to $p\sim 1/S$. This gives the required scaling properties in large-$S$ limit.}.
Other than this seemingly extreme properties required of the distribution, the second problem with this scaling is that it limits the parameter values for which the DCM solution is accurate, since the skewness of the distribution compromises the Normality assumption (see Appendix~\ref{SEC:accuracy}) unless we have $\mu^{(i)}/\sigma^{(i)} \gg 1$ .

A second variation of the model is to drop the restriction that the interaction elements be non-negative.  This obviously changes the intent of the model as, from a starting point in which all interactions are e.\,g. trophic interactions, negative values of $a_{\alpha,\beta}^{(i,j)}$ mix in some interactions that are mutualistic, some competitive, and some that remain antagonistic but opposite in direction.
Nevertheless, if the majority of the interactions remain of the original type, it still makes sense to differentiate the two guilds along the original lines. 
Such a model might be appropriate for bipartite plant–microbe networks \citep{Bennett2018} where interactions are complicated and of different signs \citep{Trivedi2020,He2021}.
Dropping this restriction does not change derivation of the DCM equations, but it does affect the validity argument presented in Appendix~\ref{SEC:accuracy}. 
If we choice the scaled matrix elements $b_{\alpha,\beta}^{(i,j)}$ from a normal distribution (which necessarily permits negative values), then we remove any concerns about the Normality of the final fluctuations, and there are no limits on the validity of the DCM equations from a skewness point of view. This means that the scaling of \eq{EQ:ABoverS} works just as well as the scaling of \eq{EQ:AB}.  However, preservation of guild interaction identity still requires $\mu^{(i)}/\sigma^{(i)} \gg 1$, and so the useful parameter regime of the model stays the same.

Looking to the future, this work opens up the study of community assembly within other block-structured ecological networks.  Here we think of two particular geometries: ``hub and spokes'' in which a central guild interacts with a number of further guild, as in \cite{Pocock2012}, and a ``ladder'' such as a food web with perfect trophic coherence, i.e. where basal species are consumed exclusively by primary consumers, primary consumers are consumed exclusively by secondary consumers, and so forth \citep{Johnson2014}.
Tripartite ecological networks \citep{Fontaine2011,Sauve2014,Sauve2016,DominguezGarcia2021,Emary2022} span both categories, and could either describe a single interaction type e.g. antagonism in a plant-pest-parasitoid network, or mixed interactions such as in plant-mutualist-parasitoid or herbivore-plant-mutualist networks.
Following the approach set out here, the DCM should allow us to map the persistence and coexistence conditions across these diverse network structures which, although certainly just caricatures, represent important aspects of the organisation of interactions central to natural ecosystems.

\begin{acknowledgements}
This work was supported by the Natural Environment Research Council (NERC) funded ONE Planet Doctoral Training Partnership (Grant Number [NE/S007512/1]).  We acknowledge helpful discussions with Darren M. Evans.
\end{acknowledgements}


\appendix
\section{Derivation of the DCM equations \label{SEC:DCM}}

We begin by defining the scaled abundances
$
  n_\alpha^{(i)} := g N_\alpha^{(i)}
$ where 
$ g =\rb{\langle N^{(1)}\rangle +\langle N^{(2)}\rangle}^{-1}$.
Then, setting $d n_\alpha^{(i)} / dt = 0$ and using \eq{EQ:AB}, the equilibrium condition of \eq{EQ:EoM1} can be written
as
\begin{align}
    0
    &= n_\alpha^{(i)}
    \left(
      \lambda_\alpha^{(i)}  
      + \xi_\alpha^{(i)} 
      - u^{(i)}n_\alpha^{(i)} 
      + \frac{1}{\sqrt{S}}c^{(i)}
      \sum_{\beta=1}^{S^{(i+1)}} b_{\alpha,\beta}^{(i,i+1)}n_\beta^{(i+1)} 
      + h^{(i)}
    \right) 
    \label{EQ:equi1}
    ,
\end{align}
in which we have defined
\begin{align}
  &u^{(i)} = \frac{1}{\sigma^{(i)}}
  ;\qquad
  f^{(i)} = g\langle N^{(i)}\rangle
  ;\qquad
  \lambda_\alpha^{(i)} 
  = 
  u^{(i)}gt^{(i)}\rb{
    K_\alpha^{(i)}
    - 
    1
  }
  ;\nonumber\\
  &~~~~~~~
  h^{(i)} = 
  t^{(i)}u^{(i)}g 
  + 
  \sqrt{S}\rho^{(i+1)}c^{(i)}\mu^{(i)}u^{(i)}f^{(i+1)}
  ,
\end{align} 
and, for purposes of derivation, added small perturbation $\xi_\alpha^{(i)}$ to each $\lambda_\alpha^{(i)}$.
The response to these perturbations can be captured by matrices $\mathbf{v}^{(i,j)}$ with elements  
\begin{align}
   v^{(i,j)}_{\alpha,\beta}:= \left[\frac{\partial n^{(i)}_{\alpha}}{\partial\xi^{(j)}_{\beta}}\right]_{0} 
   \label{EQ:vdefn}
   ,
\end{align}
evaluated at $\xi^{(i)}_\alpha = 0;~ \forall i,\alpha$.
To derive the cavity equations, we consider an initial community with equilibrium abundances $n_\alpha^{(i)}$ and to each guild we then add a species, labelled with index 0 and assumed to be statistically homogeneous with the other species in its guild.  We denote the new equilibrium abundances as $m_\alpha^{(i)}$.  These abundances will obey the same equations as
\eq{EQ:equi1} but with $m_\alpha^{(i)}$ replacing $n_\alpha^{(i)}$ and with sums extended by one. This allows us to identify the perturbations as 
\begin{align}
    \xi^{(i)}_{\alpha} = \frac{1}{\sqrt{S}}c^{(i)}b^{(i,i+1)}_{\alpha,0} m^{(i+1)}_0
    \label{EQ:xi}
    .
\end{align}
Since both $m^{(i)}_0$ and  $b^{(i,i+1)}_{\alpha,0}$ scale as $S^0$, we have $\xi_\alpha^{(i)} \rightarrow 0$ as $S\rightarrow \infty$ such that the  perturbations are weak. The abundances with and without the perturbation can therefore be related via linear response as
\begin{align}
  m^{(i)}_\alpha 
  \simeq 
  n^{(i)}_\alpha 
  +
  \sum_{j=1}^{2}
  \sum_{\beta=1}^{S^{(j)}} 
  v^{(i,j)}_{\alpha,\beta}\xi^{(j)}_\beta
  =
  n^{(i)}_\alpha 
  +
  \frac{1}{\sqrt{S}}
  \sum_{\beta=1}^{S^{(i)}} 
  v^{(i,i)}_{\alpha,\beta}
  c^{(i)}
  b^{(i,i+1)}_{\beta,0} m^{(i+1)}_0
  \label{EQ:eminen}
  ,
\end{align}
where, in the last expression, we have set the off-diagonal response blocks $v^{(i,i+1)}$ to zero, a result which will be justified later.
Under the assumption that $m_0^{(i)}>0$, substitution of the previous result into the corresponding equilibrium condition gives
\begin{align}
  0 &= \lambda^{(i)}_0 + \xi^{(i)}_0 - u^{(i)}m^{(i)}_0  + h^{(i)}   
  \nonumber\\
  &
  ~~~
  + \frac{1}{\sqrt{S}}c^{(i)}\sum_{\beta=1}^{S^{(i+1)}}b^{(i,i+1)}_{0,\beta}
  \left(
    n^{(i+1)}_\beta
    + 
    m^{(i)}_0 \frac{1}{\sqrt{S}}c^{(i+1)}  
    \sum_{\gamma=1}^{S^{(i+1)}} v^{(i+1,i+1)}_{\beta,\gamma} b^{(i+1,i)}_{\gamma,0}
  \right)
  .
\end{align}
This can be rearranged as
\begin{align}
    m^{(i)}_0 
    &= 
    \frac{
      h^{(i)} + \lambda^{(i)}_0 + \xi^{(i)}_0  
      + c^{(i)} \frac{1}{\sqrt{S}}\sum_{\alpha=1}^{S^{(i+1)}}b^{(i,i+1)}_{0,\alpha} n^{(i+1)}_\alpha
    }{
      u^{(i)}
      -
      \frac{1}{S}c^{(i)}c^{(i+1)}\sum_{\alpha,\beta=1}^{S^{(i+1)}}b^{(i,i+1)}_{0,\alpha}v^{(i+1,i+1)}_{\alpha,\beta}b^{(i+1,i)}_{\beta,0}
    }
    \label{EQ:m_before_averaging}
    .
\end{align}
As reasoned in \cite{Bunin2017}, the denominator of this expression is a finite number with negligible fluctuations and thus be replaced by its expectation value. Evaluating the expectation value of the sum, we obtain
\begin{align}
    \frac{1}{S}
    \left\langle 
    \sum_{\alpha,\beta=1}^{S^{(i+1)}}b^{(i,i+1)}_{0,\alpha}v^{(i+1,i+1)}_{\alpha,\beta}b^{(i+1,i)}_{\beta,0}
    \right\rangle 
    &= 
    \frac{\gamma}{S} \sum_{\alpha=1}^{S^{(i+1)}}v^{(i+1,i+1)}_{\alpha,\alpha}
     =\gamma \rho^{(i+1)} \langle v^{(i+1,i+1)}_{\alpha,\alpha} \rangle
    := \gamma\nu^{(i+1)}
     \label{EQ:vmess}
\end{align}
which defines the mean response coefficient, 
$
  \nu^{(i)}
$.
The denominator of \eq{EQ:m_before_averaging} can therefore be written as 
\begin{align}
    \hat{u}^{(i)} 
    := 
    \frac{1}{\sigma^{(i)}}-\gamma c^{(1)}c^{(2)}\nu^{(i+1)}
    \label{EQ:uhatdefn}
    .
\end{align}
Turning to the numerator in \eq{EQ:m_before_averaging}, we note that the quantities $\lambda^{(i)}$ have mean $\langle \lambda^{(i)}\rangle = 0$ and  variance
\begin{align}
    \mathrm{Var} \rb{\lambda^{(i)}} :=
    (\sigma_{\lambda^{(i)}})^2 
    = g^2  (u^{(i)})^2 \rb{S^{(i)}}^{-1}\rb{\kappa^{(i)}}^2
    \label{EQ:sigmasLK}
    ,
\end{align}
such that the numerator has mean $h^{(i)}$ and variance
\begin{align}
    \rb{R^{(i)}}^2 =  (\sigma_{\lambda^{(i)}_0})^2 + q^{(i+1)}
    ,
\end{align}
in which
\begin{align}
    q^{(i)} &:= \rho^{(i)}
  \langle (n^{(i)})^2 \rangle
    .
\end{align}
We have dropped the perturbation $\xi_0^{(i)}$ from these expression as, from \eq{EQ:xi}, they scale as $\mathbb{O}(S^{-1/2})$ and play no further role.
As the sum of many weakly correlated terms, we expect $c^{(i)}\sum_{\alpha=1}^{S^{(i+1)}}b^{(i+1)}_{0,\alpha}n^{(i+1)}_\alpha$ to be normally distributed in the large-$S$ limit.  Thus the numerator is equivalent to 
$
  h^{(i)} + z^{(i)} R^{(i)}
$,
where $z^{(i)}$ is distributed according to a the standard normal $z^{(i)} \sim \mathcal{N}(0,1)$.
From the Lotka-Volterra dynamics, we know that if $m^{(i)}_0>0$ then the fixed point at $m^{(i)}_0=0$ is unstable against invasion, and vice versa.  Thus, we obtain the abundance  
$
  m^{(i)}_0 = 
  \text{max}
  \left[
    0,
    \rb{h^{(i)}+z^{(i)}R^{(i)}}/{\hat{u}^{(i)}}
  \right]
$.
Once added, these additional species are no different from the other species in their guild, and thus this result holds for a general species i.e. $m^{(i)}_0 = m^{(i)}_\alpha = m^{(i)} $.  The final result is that the abundances of the species in guild $i$ are distributed according to the truncated Gaussian
\begin{align}
    m^{(i)} = \text{max}\left(
    0,
    \frac{h^{(i)}+z^{(i)}R^{(i)}}
    {\hat{u}^{(i)}}
    \right) 
    ;\qquad 
    z^{(i)} \sim \mathcal{N}(0,1)
    \label{EQ:PopnDistn}
    .
\end{align}

It now  remains to find expressions for the quantities contained in  \eq{EQ:PopnDistn}. Expectation values of function of the abundances are readily evaluated using the  probability distribution associated with \eq{EQ:PopnDistn}. We obtain
\begin{align}
    f^{(i)}&=\langle m^{(i)}\rangle 
    = 
    \frac{R^{(i)}}{\hat{u}^{(i)}}w_1(\Delta^{(i)})
    \label{EQ:fgood}
    ;
    \\
    q^{(i)}&=\rho^{(i)}(c^{(i+1)})^2\left\langle (m^{(i)})^2\right\rangle  
    = \rho^{(i)} \rb{\frac{R^{(i)}}{\hat{u}^{(i)}}}^2 w_2(\Delta^{(i)})
   \label{EQ:qgood}
   ;
    \\
    \phi^{(i)} &= \left\langle \Theta(m^{(i)})\right\rangle = w_0(\Delta^{(i)})
    ,
\end{align} 
expressed in terms of 
\begin{align}
    \Delta^{(i)} := \frac{h^{(i)}}{R^{(i)}}
    =
    \frac{t^{(i)}u^{(i)}g
    +c^{(i)}\sqrt{S}\rho^{(i+1)}\mu^{(i)}u^{(i)}f^{(i+1)}}{R^{(i)}}
    \label{EQ:Delta1}
    ,
\end{align}
and
\begin{align}
    R^{(i)}
    &=
    \sqrt{\rb{ \frac{g \kappa^{(i)}}{\sigma^{(i)} \sqrt{S^{(i)}}} }^2 + q^{(i+1)}}
    \label{EQ:Rdefn}
    .
\end{align}
Expressions for $\nu^{(i)}$ can be obtained by differentiating \eq{EQ:m_before_averaging} and noting that we only obtain a non-zero contribution with a probability $\phi^{(i)}$: 
\begin{align}
    \nu^{(i)} 
    = \rho^{(i)}\langle v^{(i,i)}_{0,0}\rangle  
    = \rho^{(i)}\left\langle\left[\frac{\partial m^{(i)}_0}{\partial\xi^{(i)}_0}\right]_{0}\right\rangle_{m_0^{(i)} > 0}
    = \frac{\rho^{(i)}\phi^{(i)}}{\hat{u}^{(i)}}
    \label{EQ:nu}
\end{align}
We also see from \eq{EQ:m_before_averaging} that $\partial m^{(i)}_0/\partial\xi^{(i+1)}_0 = 0$, consistent with the earlier assertion that $v_{\alpha,\beta}^{(i,i+1)} = 0$.  The final equation required to close the set is 
$
  f^{(1)}+f^{(2)}=1
$ by definition.

\section{Solution of DCM equations \label{SEC:Asol}}
 
We can collapse the above set of equations into a pair of equations, solution of which gives parameters $\Delta^{(i)}$ as a function of model parameters. 
First we use \eq{EQ:nu} to write \eq{EQ:uhatdefn} as
\begin{align}
    \hat{u}^{(i+1)}
    =   
    u^{(i+1)} - 
    \frac{\gamma c^{(1)} c^{(2)}\rho^{(i)}\phi^{(i)}}{\hat{u}^{(i)}}
    .
\end{align}
This maps onto a pair of independent quadratics with solution
\begin{align}
    \hat{u}^{(i)}
    &=  
    \frac{1}{2u^{(i+1)}}
    \left\{
      u^{(i)}u^{(i+1)}
      -\gamma c^{(1)}c^{(2)}
      \left[\rho^{(i+1)} \phi^{(i+1)} - \rho^{(i)}\phi^{(i)}\right]
    \right.
    \nonumber\\
    & 
    \left.
      +
      \sqrt{
        \left\{
            u^{(i)}u^{(i+1)}
            -\gamma c^{(1)}c^{(2)}
            \left[\rho^{(i+1)} \phi^{(i+1)} - \rho^{(i)}\phi^{(i)}\right]
        \right\}^2
        - 4 \gamma c^{(1)}c^{(2)} \rho^{(i)} \phi^{(i)} u^{(i)}
      }
    \right\}
    \label{EQ:uhatexplicit}
    ,
\end{align}
where have taken the positive root so that $\hat{u}^{(i)} \to u^{(i)}$ for $\gamma \to 0$.

We then turn to \eq{EQ:qgood} and use \eq{EQ:Rdefn} to obtain
\begin{align}
    M
   \begin{bmatrix}
       \rb{R^{(1)}}^2 \\ \rb{R^{(2)}}^2 
   \end{bmatrix}
   =
   \frac{g^2}{S} 
   \begin{bmatrix}
       \rb{u^{(1)}\kappa^{(1)}}^2 / \rho^{(1)}
       \\ 
       \rb{u^{(2)}\kappa^{(2)}}^2 / \rho^{(2)}
   \end{bmatrix}   
   \label{EQ:Meqn}
   ,
\end{align}
with the matrix
\begin{align}
    M = 
   \begin{bmatrix}
       1 & 
       - \rho^{(2)} 
       w_2\rb{\Delta^{(2)}} / \rb{\hat{u}^{(2)}}^2 \\
       - \rho^{(1)}
       w_2\rb{\Delta^{(1)}} / \rb{\hat{u}^{(1)}}^2
       & 1
   \end{bmatrix}
   \label{EQ:M}
   .
\end{align}
There are then two different solution routes, depending on whether both $\kappa^{(i)} =0$ or not. Assuming at least one $\kappa^{(i)}$ is non-zero and that $M$ is non-singular, we obtain a solution 
$
    R^{(i)} = g S^{-1/2}P^{(i)}
$
with
\begin{align}
   \begin{bmatrix}
       \rb{P^{(1)}}^2 \\ \rb{P^{(2)}}^2 
   \end{bmatrix}
   = M^{-1}
   \begin{bmatrix}
       \rb{u^{(1)}\kappa^{(1)}}^2 /\rho^{(1)}
       \\ 
       \rb{u^{(2)}\kappa^{(2)}}^2 /\rho^{(2)}
   \end{bmatrix}
   ,
\end{align}
where we take the positive root $P^{(i)}>0$ since $R^{(i)}>0$. 
This gives
\begin{align}
    f^{(i)} 
    = 
    g S^{-1/2} \frac{P^{(i)}}{\hat{u}^{(i)}}w_1(\Delta^{(i)})
    ,
\end{align}
such that \eq{EQ:Delta1} may be written as
\begin{align}
    1 &= 
    \frac{
        \Delta^{(i)} S^{-1/2} P^{(i)}\hat{u}^{(i+1)}
        -
        c^{(i)} \mu^{(i)}u^{(i)} 
        w_1(\Delta^{(i+1)})
        \rho^{(i+1)}
        P^{(i+1)}
    }{
        t^{(i)}u^{(i)}\hat{u}^{(i+1)}
    } 
    .
\end{align}
The two resultant equations can then be solved to give $\Delta^{(i)}$ as functions of model parameters. The normalisation constraint $f^{(1)} + f^{(2)} =1$ then determines an expression for $g$, namely
\begin{align}
    g &= 
    \frac{
        \hat{u}^{(1)} \hat{u}^{(2)}S^{1/2}
    }{
         \hat{u}^{(2)}  P^{(1)} w_1 (\Delta^{(1)})
         +
         \hat{u}^{(1)}  P^{(2)} w_1 (\Delta^{(2)})
    }
    \label{EQ:gexpressionkappane0}
    .
\end{align}

In the case when $\kappa^{(1)} = \kappa^{(2)} = 0$, the righthand side of \eq{EQ:Meqn} vanishes.  Thus we have
\begin{align}
    \rb{R^{(1)}}^2 
    &= 
    \rho^{(2)} \rb{\frac{1}{\hat{u}^{(2)}}}^2 
    w_2(\Delta^{(2)}) \rb{R^{(2)}}^2
    \label{EQ:R1R2sing}
    .
\end{align}
Matrix $M$ must then be singular, which yields
\begin{align}
    \rb{\hat{u}^{(1)} \hat{u}^{(2)}}^2
    & =
    \rho^{(1)} \rho^{(2)} w_2(\Delta^{(1)})w_2(\Delta^{(2)})
    \label{EQ:singeqn1}
    .
\end{align}
We can then use \eq{EQ:fgood} and \eq{EQ:R1R2sing} as well as both instances of \eq{EQ:Delta1} to obtain
\begin{align}
    \frac{g}{R^{(2)}} 
    &=
    \frac{\Delta^{(1)}}{t^{(1)} u^{(1)}\hat{u}^{(2)}}\sqrt{\rho^{(2)} w_2 (\Delta^{(2)})}
    -
    \frac{c^{(1)} \sqrt{S}\rho^{(2)}\mu^{(1)} w_1(\Delta^{(2)})}{t^{(1)} \hat{u}^{(2)}}
    \nonumber\\
    &=    
    \frac{\Delta^{(2)}}{t^{(2)} u^{(2)}}
    -
    \frac{c^{(2)}\rho^{(1)} \mu^{(2)} w_1(\Delta^{(1)}) }{t^{(2)} \hat{u}^{(1)}\hat{u}^{(2)}}
    \sqrt{S^{(2)} w_2 (\Delta^{(2)})}
    \label{EQ:singeqn2}
    .
\end{align}
In this case, \eq{EQ:singeqn1} and  \eq{EQ:singeqn2} are the pair of equations that connect $\Delta^{(i)}$ and model parameters. We then use the normalisation $1 = f^{(1)} + f^{(2)}$ to obtain explicit forms for $R^{(i)}$.  We find
\begin{align}
    R^{(2)}
    &=
    \frac{
        \hat{u}^{(1)} \hat{u}^{(2)}
    }{
      w_1\rb{\Delta^{(2)}} \hat{u}^{(1)}
      +
      w_1\rb{\Delta^{(1)}}\sqrt{w_2\rb{\Delta^{(2)}}}
    }
    ,
\end{align}
from which
\begin{align}
    f^{(2)}
    &=
    \frac{
        w_1\rb{\Delta^{(2)}} \hat{u}^{(1)}
    }{
      w_1\rb{\Delta^{(2)}} \hat{u}^{(1)}
      +
      w_1\rb{\Delta^{(1)}}\sqrt{w_2\rb{\Delta^{(2)}}}
    }
    ,
\end{align}
and
$f^{(1)} = 1-f^{(2)}$.
This gives the ratio of the total abundances in equilibrium as
\begin{align}
  \mathcal{R} 
  :=
  \frac{\sum_{\alpha=1}^{S^{(2)}} N ^{(2)}_\alpha}
  {\sum_{\alpha=1}^{S^{(2)}} N ^{(1)}_\alpha}
  = 
  \frac{\rho^{(2)} f^{(2)}}{\rho^{(1)} f^{(1)}}
  =
  \frac{ \rho^{(2)} }{\rho^{(1)}}
  \frac{
   \hat{u}^{(1)}w_1\rb{\Delta^{(2)} }
  }{
    w_1\rb{\Delta^{(1)}}\sqrt{w_2\rb{\Delta^{(2)}}}
  }
  \label{EQ:Rratiodefn}
  .
\end{align}

\section{Validity of the DCM analysis \label{SEC:accuracy}}

The DCM solution involves replacing the numerator in  \eq{EQ:m_before_averaging} with a normally-distributed random variable. 
The Berry-Esseen theorem \citep{Berry1941,Esseen1942} supplies an upper bound for the error in the cumulative-density function approximated in this way. Applying this to the case in hand, with $\kappa^{(i)}=0$ initially, we find that error bound to be
\begin{align}
  \mathcal{E}^{(i)}
  = 
 \rb{ C
  \frac{\ew{(n^{(i+1)})^3}}{\ew{(n^{(i+1)})^2}^{3/2}}
  }
  \cdot
  \rb{
   \frac{1}{\sqrt{\rho^{(i+1)} S}}
  \ew{\left|b^{(i,i+1)}_{0,\alpha}\right|^3}
  }
  \label{EQ:epserror}
  ,
\end{align}
where $C$ is some constant.
Since the first factor only contains moments of the fractional variables $n^{(i+1)}_\alpha$, it scales like $S^0$.  This leaves the scaling behaviour of the error term dependent on the second factor.  To calculate $\ew{\left|b^{(i,i+1)}_{0,\alpha}\right|^3}$ we need to go back to \eq{EQ:AB} and consider the original distribution of the matrix elements  $a^{(i,i+1)}_{0,\alpha}$. For a half-normal distribution for example, we find that 
$\ew{\left|b^{(i,i+1)}_{0,\alpha}\right|^3}\approx 1.72$, i.e. a constant.  For a gamma distribution we find 
\begin{align}
    \ew{\left|b^{(i,i+1)}_{0,\alpha}\right|^3}
    \sim \frac{2\sigma^{(i)}}{\mu^{(i)}}
    .
\end{align}
The key point is that in both cases this value is independent of $S$. The result is that  $\mathcal{E}^{(i)}$ vanishes in the limit thanks to the explicit $S^{-1/2}$ dependence of \eq{EQ:epserror}. It seems plausible that this behaviour extends to any similar distribution with the same scaling.

For $\kappa^{(i)} \ne 0$, the fluctuations in $\lambda$ also need to be taken into account. Considering these independently of the interactions and taking the distribution of $K^{(i)}$ to a gamma distribution, the error associated with the normal approximation will be proportional to 
\begin{align}
  \frac{\kappa^{(i)}}{
  \sqrt{S}
  }
  ,
\end{align}
Thus, with matrix elements as in \eq{EQ:AB}, the error associated with the normality approximation scales like $S^{-1/2}$. This is consistent with other approximations in the derivation and means that in the $S\to \infty$ limit, the DCM solution becomes exact.

At finite $S$, the above limitations do have a bearing on the parameters for which we can  expect agreement between numerics and DCM expressions. With distributions chosen as in \secref{SEC:numerics}, close agreement requires $\sqrt{S}\gg 1$ and $\sqrt{S}\gg\kappa^{(i)}$. The main effect of this is the restriction of the value of $\kappa^{(i)}$ used in the plots here to  $\kappa^{(i)} \le 5$.

In the alternative scaling scheme discussed in \eq{EQ:ABoverS} with $a^{(i,i+1)}_{0,\alpha}>0$, we obtain $\ew{\left|b^{(i,i+1)}_{0,\alpha}\right|^3} \sim \sqrt{S}$ and therefore the overall error scales as $S^0$. In this case we can no longer rely on the asymptotic limit to make our distributions converge to the Gaussian limit.  Rather, the accuracy of the DCM solutions is dependent on other model parameters. In particular, for interaction coefficients distributed according to a Gamma distribution, we would require 
$
\sigma^{(i)}\sqrt{\rho^{(i+1)}} / \mu^{(i)}\ll 1
$.  Lifting the $a^{(i,i+1)}_{0,\alpha}>0$ restriction changes this requirement. For example, we might then choose the elements themselves to be normally distributed, in which case $\ew{\left|b^{(i,i+1)}_{0,\alpha}\right|^3}=0$ and any potential convergence issues vanish (in either scaling approach).

\section{Stability \label{SEC:stability}}

The unbounded phase occurs when one or both $\ew{N^{(i)}} \to \infty$, which means that we can find the phase boundary in the DCM analysis by considering $g \to 0$.
Looking first at the $\kappa^{(i)} \ne 0$ case, we see that $g \to 0$ in \eq{EQ:gexpressionkappane0} implies that at least one of the $P^{(i)}$ diverges.  This occurs when the matrix $M$ of \eq{EQ:M} is singular. This is exactly the situation discussed for $\kappa^{(i)} = 0$ and thus the phase boundary for arbitrary $\kappa^{(i)} $ can be obtained from the equations of the $\kappa^{(i)} = 0$ case.

At the phase boundary, \eq{EQ:singeqn1} still holds, as does 
\eq{EQ:singeqn2} but now with the addition that both sides of the equality are individually zero (from $g =0$).  Using \eq{EQ:singeqn1}, these latter equations can be written as
\beq
  \Delta^{(i)}\frac{\sqrt{w_2(\Delta^{(i+1)})}}{w_1(\Delta^{(i+1)})}
  &=& 
   c^{(i)}
   \frac{\sqrt{S^{(i+1)}} \mu^{(i)}}{\sigma^{(i)}}
  \label{EQ:DeltaPTN1_i}
  .
\eeq
A plot of the function $Q(\Delta) := \sqrt{w_2(\Delta)}/w_1(\Delta)$ shows it to be a monotonically-decreasing function that diverges exponentially for $\Delta \to -\infty$ and has a limit $Q(\Delta)\to 1$ for $\Delta \to \infty$; $Q(\Delta)$. If has a value at the origin of $Q(0)= \sqrt{\pi}$.  Using these properties, we see that \eq{EQ:DeltaPTN1_i} implies that $\mathrm{sign}\,[\Delta^{(i)}] = c^{(i)}$. In the antagonistic case, we have $c^{(1)} = - c^{(2)} = -1$ such that  $\Delta^{(2)} > 0$. This means that $Q(\Delta^{(2)}) < \sqrt{\pi}$ and therefore from  \eq{EQ:DeltaPTN1_i} we have
\beq
  \Delta^{(1)} < - \frac{\sqrt{S^{(2)}} \mu^{(1)}}{\sqrt{\pi}\sigma^{(1)}}
  \label{EQ:D1>}
  .
\eeq
For large $S$, this result implies that $Q(\Delta^{(1)})$ is exponentially large which, from \eq{EQ:DeltaPTN1_i}, means that $\Delta^{(2)}$ will be exponentially small.  Setting $\Delta^{(2)} \to 0$ into \eq{EQ:DeltaPTN1_i} turns the inequality of \eq{EQ:D1>} into an equality.
Thus in the limit the $\Delta^{(i)}$-values at the boundary become
\beq
  \Delta^{(1)}  = - \frac{\sqrt{S^{(2)}} \mu^{(1)}}{\sqrt{\pi}\sigma^{(1)}} 
  ;\qquad 
   \Delta^{(2)}  = 0
   \label{EQ:DeltaPTNsolved}
   .
\eeq
Substituting these values into \eq{EQ:singeqn1} gives us the location of the phase boundary as being determined by
\begin{align}
    \rb{\hat{u}^{(1)} \hat{u}^{(2)}}^2
    & =
    \frac{1}{2}\rho^{(1)} \rho^{(2)} w_2\rb{- \frac{\sqrt{S^{(2)}} \mu^{(1)}}{\sqrt{\pi}\sigma^{(1)}} }
    .
\end{align}
Here $\hat{u}^{(i)}$ are evaluated at the values of \eq{EQ:DeltaPTNsolved} and are thus  functions of the model parameters only.

\section{Transition to multiple attractors \label{SEC:MA}}

The stability of the cavity solution can be studied with a generalisation of the method presented in the appendix of \cite{Bunin2017} to the structure relevant here.  Our starting point in the expression
\begin{align}
    m^{(i+)}_0 &= 
    \frac{
      h^{(i)}+\lambda^{(i)}_0+\xi^{(i)}_0  + c^{(i)}S^{-1/2}\sum_{\alpha=1}^{S^{(i+1)}}b^{(i,i+1)}_{0,\alpha}n^{(i+1)}_\alpha
    }{
      \hat{u}^{(i)}
    }
    \label{EQ:m_before_averaging2}
    ,
\end{align}
which is \eq{EQ:m_before_averaging} with the denominator evaluated as previously discussed, and with the ``+'' added to remind us that this is the abundance when positive.

We consider a vector of perturbation $\bs{\xi}^{(i)} = \epsilon \bs{\eta}^{(i)}$ with mean $ \overline{\bs{\eta}^{(i)} }=0 $. The derivative of \eq{EQ:m_before_averaging2} with respect to perturbation ``strength'' $\epsilon$ is
\begin{align}
    y_0^{(i+)} 
    := 
    \frac{d m_0^{(i+)}}{d\epsilon} 
    =
    \frac{1}{\hat{u}^{(i)}}
    \rb{
     \eta^{(i)}_0  
     + c^{(i)} S^{-1/2} \sum_{\alpha=1}^{S^{(i+1)}}b^{(i,i+1)}_{0,\alpha}y^{(i+1)}_{\alpha/0}
    }
    \label{EQ:y1}
    ,
\end{align}
in which
$
    y_{\alpha/0}^{(i)} := \frac{d n_\alpha ^{(i)}}{d\epsilon} 
$. Note that when $m_0>0$, we have $\frac{d m_\alpha ^{(i)}}{d\epsilon} =  \frac{d m_\alpha ^{(i+)}}{d\epsilon}$, otherwise $\frac{d m_\alpha ^{(i)}}{d\epsilon} = 0 $.  Squaring \eq{EQ:y1} we have
\begin{align}
    \rb{\hat{u}^{(i)}}^2 \rb{y_{0}^{(i+)} }^2
    &=
    \rb{\eta^{(i)}_0 }^2
    + 
    2\eta^{(i)}_0  c^{(i)}S^{-1/2}\sum_{\alpha=1}^{S^{(i+1)}}b^{(i,i+1)}_{0,\alpha}y^{(i+1)}_{\alpha/0}
    \nonumber\\
    &
    ~~~~
    + 
    S^{-1}\sum_{\alpha,\alpha'=1}^{S^{(i+1)}}
        b^{(i,i+1)}_{0,\alpha}b^{(i,i+1)}_{0,\alpha'} 
        y^{(i+1)}_{\alpha/0}y^{(i+1)}_{\alpha'/0}
    \nonumber
    .
\end{align}
The final term is self averaging, such that 
\begin{align}
    \rb{\hat{u}^{(i)}}^2 \rb{y_{0}^{(i+)} }^2
    &=
    \rb{\eta^{(i)}_0 }^2
    + 
    2\eta^{(i)}_0  c^{(i)} S^{-1/2}\sum_{\alpha=1}^{S^{(i+1)}}b^{(i,i+1)}_{0,\alpha}y^{(i+1)}_{\alpha/0}
    \nonumber\\
    &~~~
    + 
     \rho^{(i+1)} \phi^{(i+1)}
    \ew{\rb{ y^{(i+1)}_{\alpha/0}}^2}_+
    ,
\end{align}
in which $\ew{\ldots}_+$ denotes the average over positive abundances only.
Averaging over $\eta_0^{(i)}$, which is independent from $b^{(i,i+1)}_{0,\alpha}$ and $y^{(i+1)}_{\alpha/0}$, we obtain
\begin{align}
    \rb{\hat{u}^{(i)}}^2 \overline{\rb{y_{0}^{(i+)} }^2}
    &    =
    \overline{\rb{\eta^{(i)}_0 }^2}
    + 
    \rho^{(i+1)} \phi^{(i+1)}
    \ew{\rb{ y^{(i+1)}_{\alpha/0}}^2}_+
    .
\end{align}
Thus $\overline{\rb{y_{0}^{(i+)} }^2}$ is independent of $n_0^{(i)}$.  Once added to a guild, species $0$ is the same as any other, so that $\overline{\rb{y_{0}^{(i+)} }^2} =  \ew{\rb{ y^{(i)}_{\alpha/0}}^2}_+ =  \ew{\rb{ y^{(i)}_{\alpha}}^2}_+$.  Thus we obtain 
\begin{align}
    \rb{\hat{u}^{(i)}}^2 \ew{\rb{ y^{(i)}_{\alpha/0}}^2}_+
    &    =
    \overline{\rb{\eta^{(i)}_0 }^2}
    + 
    \rho^{(i+1)} \phi^{(i+1)}
    \ew{\rb{ y^{(i+1)}_{\alpha/0}}^2}_+
    .
\end{align}
Writing out the two equations and arranging into matrix form, we have
\begin{align}
    \begin{bmatrix}
      \rb{\hat{u}^{(1)}}^2 & -  \rho^{(2)} \phi^{(2)}\\
      -  \rho^{(1)} \phi^{(1)} & \rb{\hat{u}^{(2)}}^2 \\
    \end{bmatrix}
    \begin{bmatrix}
      \ew{\rb{ y^{(1)}_{\alpha/0}}^2}_+ \\ \ew{\rb{ y^{(2)}_{\alpha/0}}^2}_+
    \end{bmatrix}
    &  =
    \begin{bmatrix}
      \overline{\rb{\eta^{(1)}_0 }^2} \\ \overline{\rb{\eta^{(2)}_0 }^2}
    \end{bmatrix}
    .
\end{align}
Solution for $ \ew{\rb{ y^{(i)}_{\alpha/0}}^2}_+$ requires that the above $2\times2$ matrix be invertible. Non-invertibility is taken as the signal of the UFP-MA phase transition, which occurs when
\begin{align}
    \mathrm{det}
    \begin{bmatrix}
      \rb{\hat{u}^{(1)}}^2 & -  \rho^{(2)} \phi^{(2)}\\
      -  \rho^{(1)} \phi^{(1)} & \rb{\hat{u}^{(2)}}^2 \\
    \end{bmatrix}
   =0.
\end{align}
Thus the  UFP-MA phase boundary is given by
\begin{align}
  \phi^{(1)}\phi^{(2)} 
  =
  w_0(\Delta^{(1)}) w_0(\Delta^{(2)})
  =
  \frac{\rb{\hat{u}^{(1)}\hat{u}^{(2)}}^2}
  {\rho^{(1)} \rho^{(2)}}
    .
\end{align}

\section{Asymptotic limit \label{SEC:Aasym}}

For $f^{(i)}$ to be bounded $0\le f^{(i)} \le 1$, quantities $\Delta^{(i)}$ have to scale like $S^0$. Thus taking the $S \to \infty$  limit of \eq{EQ:Delta1} yields
\begin{align}
    g &=\frac{ -c^{(i)} \sqrt{S}\rho^{(i+1)}\mu^{(i)}u^{(i)}f^{(i+1)}}{t^{(i)}u^{(i)}} ; \quad i = 1,2
    .
\end{align}
Equating the two instances and using $f^{(1)}+f^{(2)} = 1$,  we have the expressions 
\begin{align}
    f^{(i)} &= 
    \frac{
      t^{(i+1)} c^{(i)} \mu^{(i)} \rho^{(i+1)}
     }{
      t^{(2)} c^{(1)} \mu^{(1)} \rho^{(2)}
      + 
      t^{(1)} c^{(2)} \mu^{(2)} \rho^{(1)}
    }  
    \nonumber\\
    g &=
    \frac{
      -c^{(1)}c^{(2)} \mu^{(1)}\mu^{(2)} \rho^{(1)} \rho^{(2)} \sqrt{S}
    }{
      t^{(2)} c^{(1)} \mu^{(1)} \rho^{(2)}
      + 
      t^{(1)} c^{(2)} \mu^{(2)} \rho^{(1)}
    }
    .
\end{align}
These we obtain, therefore, independent of parameters $\Delta^{(i)}$.
As a consequence, the ratio of total abundances of the two guilds of \eq{EQ:Rratiodefn}
simply evaluates as
\begin{align}
  \mathcal{R} 
   =
  \frac{
     \mu^{(2)} 
    }{
     \mu^{(1)} 
    }
  .
\end{align}
The equations for $\Delta^{(i)}$ also simplify in this limit, but the resulting forms in general deliver little further insight. An exception to this is the $\kappa=\gamma =0$ case for which \eq{EQ:singeqn2} can be written as 
\begin{align}
    \rb{\frac{w_1(\Delta^{(1)})}{w_1(\Delta^{(2})}}^2
    \sqrt{\frac{w_2(\Delta^{(2)})}{w_2(\Delta^{(1})}}
    &=
    \frac{\rho^{(2)}}{\rho^{(1)}}
    \rb{\frac{\mu^{(1)}}{\mu^{(2)}}}^2
    \frac{u^{(1)}}{u^{(2)}}
    .
\end{align}
For symmetric parameters ($\rho^{(i)} = \rho = 1/2$, $\sigma^{(i)} = \sigma$ etc)
the righthand side evaluates to one.
The behaviour of the $w_i$ functions is such that this is only fulfilled when $\Delta^{(1)} = \Delta^{(2)} = \Delta$ and thus the abundance distribution of the two guilds becomes identical. It then follows from \eq{EQ:singeqn2main} that the equation for $\Delta$ becomes
$
    w_2(\Delta) = 2/\sigma^2
$.

\section{Numerical Simulations \label{SEC:numerics}}

In our numerical simulations, interaction coefficients $a^{(i,i+1)}_{\alpha,\beta}$were sampled from a half normal distribution with parameters satisfying: $\langle a^{(i,i+1)}_{\alpha,\beta} \rangle = \mu^{(i)}/\sqrt{S}$ and $\mathrm{Var}(a^{(i,i+1)}_{\alpha,\beta}) = \rb{\sigma^{(i)}}^2/S$. Use of the half-normal distribution means that $\sigma^{(i)}$ and $\mu^{(i)}$ are related as $\mu^{(i)}= \sqrt{2/(\pi-2)} \sigma^{(i)}$.
We then combine the $a^{(i,i+1)}_{\alpha,\beta}$ terms into the matrix $\mathbf{A}$, including correlation between the sub-matrices by setting $a^{(1,2)}_{\alpha,\beta} = \gamma a^{(2,1)}_{\beta,\alpha} + \sqrt{1-\gamma^2}a^{(1,2)}_{\alpha,\beta} $. As $\gamma \in [0,1]$ this choice preserves the mean and standard deviation of the two sub-matrices (assuming they both have the same distribution,as here). 
The carrying capacities $K^{(i)}$ were sampled from a gamma distribution with parameters such that $\langle K^{(i)} \rangle = 1$ and $\mathrm{Var} \rb{K^{(i)}} = \rb{S^{(i)} }^{-1}\rb{\kappa^{(i)}}^2  $.
Since we are interested in the equilibrium properties, the details of the rates $r^{(i)}$ are unimportant and for convenience, we set them as  $r^{(i)}_\alpha = K^{(i)}_\alpha$. This sets the forefactors of \eq{EQ:EoM1} equal to one and avoids adding extra time scales into the problem when $K^{(i)}_\alpha$ are drawn from a distribution.

Initial conditions for the abundances  $N^{(i)}_\alpha$ were drawn from a uniform distribution on $[0,\frac{1}{\sqrt{S}}]$.  The Lotka-Volterra dynamics of \eq{EQ:EoM1} were then integrated numerically using a Runge-Kutte-4 method. When the abundance of a species reached a threshold of $10^{-9}/\sqrt{S}$, it was set to zero and treated as being extinct in the local community. The $1\sqrt{S}$ scaling of these quantities was adopted because of the analytic results such as \eq{EQ:gexpressionkappane0} suggesting that typical equilibrium abundances scale like $1\sqrt{S}$, a fact born out by simulation.
The solver was then terminated either when an approximate equilibrium was found, defined as when all $\frac{dN^{(i)}_\alpha}{dt}<10^{-11}/\sqrt{S}$, or the maximum time of $t=10^5$ was reached. 
This process is then repeated for a total of 25 times in order to take the average.

%

\end{document}